\documentclass[USenglish, onecolumn]{scrartcl}
\usepackage[USenglish]{babel}

\usepackage{letltxmacro}
\LetLtxMacro{\ORIGselectlanguage}{\selectlanguage}
\makeatletter
\DeclareRobustCommand{\selectlanguage}[1]{%
  \@ifundefined{alias@\string#1}
    {\ORIGselectlanguage{#1}}
    {\begingroup\edef\x{\endgroup
       \noexpand\ORIGselectlanguage{\@nameuse{alias@#1}}}\x}%
}
\newcommand{\definelanguagealias}[2]{%
  \@namedef{alias@#1}{#2}%
}
\makeatother
\definelanguagealias{en}{english}

\usepackage[numbers]{natbib}

\usepackage[utf8]{inputenc}
\usepackage{lmodern}
\usepackage{microtype}

\usepackage{authblk}

\usepackage{abstract}
\newcommand{\placeabstract}[1]{\begin{abstract}\noindent{}#1\end{abstract}}

\newcommand{\figpath}{./}
\newcommand{\graphpath}{./}

\usepackage{amsmath}
\usepackage{amssymb}
\usepackage{nicefrac}
\usepackage{mathtools}
\usepackage{siunitx}
\usepackage{hyperref}

\usepackage{graphicx}

\begin{document}

\title{Dinosaur Photonic Crystal Cavity Interfaces for Color Center Coupling to Triangular Nanostructures}

\author[1,2]{Julian~M.~Bopp}
\author[1,2]{Lucca Valerius}
\author[1,2,*]{Tim~Schröder} 
\affil[1]{Department of Physics, Humboldt-Universität zu Berlin, Newtonstr. 15, 12489 Berlin, Germany}
\affil[2]{Ferdinand-Braun-Institut gGmbH, Leibniz-Institut für Höchstfrequenztechnik, Gustav-Kirchhoff-Str. 4, 12489 Berlin, Germany}
\affil[*]{Corresponding author: Tim Schröder, tim.schroeder@physik.hu-berlin.de}	

\maketitle

\placeabstract{%
Waveguide--coupled photonic crystal cavities with a triangular cross section fabricated by angled etching are suitable to interface embedded color centers with flying photonic qubits in quantum information applications.
Moreover, their fabrication requires fewer processing steps compared to nanostructures produced by quasi-isotropic undercutting.
As an alternative to established hole-based photonic crystal cavities, we introduce corrugated triangular `Dinosaur' photonic crystal cavities, and develop a tapered, quasi loss-free cavity--waveguide interface to adiabatically interconvert Bloch and waveguide modes.
We optimize the cavity--waveguide interface to minimize photon losses and demonstrate that its adjustment allows precise tuning of the light-matter interaction.
}

\section{Introduction}
\begin{figure}[htbp]
	\centering
	\includegraphics[width=10.5cm]{\graphpath 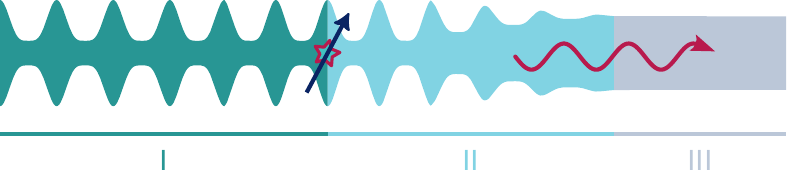}
	\caption{Concept of a Dinosaur spin--photon interface.
	A spin-active solid-state color center (star with dark blue arrow) is embedded in the center of a tapered `Dinosaur' photonic crystal cavity.
	On its left, the cavity possesses a strong mirror (section I).
	On its right, there is a weak mirror (section II) that transitions to a cavity-attached waveguide (section III) to adiabatically transfer photons (red arrow) from the cavity to the waveguide and vice versa.}
	\label{fig:concept}
\end{figure}

Spin--photon interfaces are one of the fundamental ingredients of quantum repeaters within quantum networks with distant network nodes that distribute entanglement over optical fibers \cite{Kimble2008, Wehner2018}.
Such networks providing entangled states shared across all nodes can be employed in various applications, like quantum cryptography \cite{Ekert1991} or distributed sensing \cite{Guo2019}.
Repeater nodes within these networks compensate for intrinsic fiber losses and may perform error correction on the transmitted information \cite{Muralidharan2016}.
Both involves entanglement swapping or quantum teleportation operations \cite{Briegel1998, Muralidharan2016}, which are of probabilistic nature \cite{Tittel2010}.
Transferring photonic qubits via an efficient spin--photon interface, for instance, to the spin degree of freedom of a solid-state color center (Fig.~\ref{fig:concept}) allows a quantum repeater to store received quantum information until required probabilistic operations succeed, such that these operations do not need to succeed at once in the entire network for establishing a common entangled state.
In turn, the entanglement rates of quantum networks directly depend on the efficiency of transferring a photonic to a spin qubit and vice versa \cite{Borregaard2020}, motivating the search for efficient nanophotonic spin--photon interfaces.

Considering scalable solid-state material platforms \cite{Li2024a}, photonic crystal (PhC) cavities with embedded color centers and a tapered section interfacing a cavity with a waveguide are widely employed \cite{Nguyen2019, Knall2022a}.
Usually, the PhC cavities are formed by `drilling' round, elliptical, or rectangular holes into a suspended nanobeam \cite{Mouradian2017, Knall2022a, Li2015b}.
However, nanofabrication limits the minimal achievable hole diameter and thus prevents decreasing the hole size adiabatically within tapered sections to interconvert Bloch and waveguide modes with near-unity efficiency \cite{Palamaru2001}.
Replacing holes by corrugation features along the outer nanobeam edges, we have recently demonstrated the `Sawfish' PhC cavity, which strongly suppresses scattering losses at its cavity--waveguide interface by smoothly decaying corrugation amplitudes \cite{Bopp2023a}.
As the suspended Sawfish cavity possesses a rectangular cross section fabricated by quasi-isotropic undercutting \cite{Pregnolato2024}, here we extend this cavity geometry to `Dinosaur' PhC cavities with a triangular cross section that can be realized by means of Faraday etching \cite{Burek2012}.
Faraday etching involves less steps, i.e., conformal coating for sidewall protection followed by quasi-isotropic etching is not necessary, simplifying the fabrication of freestanding, one-dimensional nanostructures.

Triangular nanostructures have been investigated for different material platforms, which may host spin-active color centers.
In diamond, respective hole-based PhC cavities have been designed and fabricated for the zero-phonon line (ZPL) of the negatively-charged nitrogen-vacancy (NV) center at a wavelength of $637\,$nm \cite{Bayn2011, Bayn2014, Burek2014} and the silicon-vacancy (SiV) center at a wavelength of $737\,$nm \cite{Zhang2018}.
Furthermore, a hole-based PhC reflector has been proposed for the NV center in 4\textit{H} silicon carbide (4\textit{H}-SiC) \cite{Majety2023}, and a PhC cavity for various 4\textit{H}-SiC color centers \cite{Majety2021}.
Color centers in both materials are deemed promising for quantum networking.
Particularly, the negatively-charged diamond NV center exhibits long spin coherence times \cite{Abobeih2018}, while the tin-vacancy (SnV) center is inversion-symmetric and can be operated at elevated temperatures compared to lighter group--IV centers due to a relatively large energy splitting between the orbital branches separating possible qubit energy levels \cite{Iwasaki2017}.
In contrast, 4\textit{H}-SiC is compatible with integration into complementary metal-oxide-semiconductor (CMOS) platforms \cite{Lukin2019}.
Moreover, the spin of its silicon-vacancy center at a \textit{k} lattice site (\textit{k}-$\mathrm{V}_\mathrm{Si}$) can be manipulated at room temperature \cite{Widmann2015}.

Taking advantage of corrugation-based photonic crystal cavities, in this work we develop Dinosaur PhC cavities, which we first design and simulate for the \textit{k}-$\mathrm{V}_\mathrm{Si}$ center in 4\textit{H}-SiC adjusted to its V2 ZPL frequency of $326.9\,$THz ($917.1\,$nm) \cite{Soerman2000}.
Next, we introduce a tapered cavity--waveguide interface and optimize the cavity--waveguide coupling efficiency, before adapting the Dinosaur cavity to diamond color centers.

\section{Simulation Methods}
We use the software package \textit{JCMsuite} \cite{JCMsuite} to determine eigenmodes of periodic or resonant PhC nanostructures in finite element method (FEM) simulations.
Cavity mode quality factors $Q$ are derived from the simulated complex eigenfrequencies.
Integrating the simulated electric field energies over the entire computational domain yields the corresponding mode volumes $V$.
Together, both determine the Purcell factor $F_\mathrm{P}\propto Q/V$ of a cavity mode \cite{Purcell1946}.
Scattering simulations, where a dipole radiating at the cavity's fundamental eigenfrequency is positioned in the center of a waveguide-attached cavity, reveal the cavity--waveguide coupling efficiency $\beta_\mathrm{WG}$.
This efficiency is the fraction of the dipole's total emitted power that results from integrating the Poynting vector across the computational domain's delimiting plane the waveguide intersects with.

Adapting Dinosaur PhC cavities to the ZPL frequencies of distinct color centers involves geometry parameter sweeps first.
Second, the quality factor of the fundamental cavity mode is maximized with JCMsuite's integrated Bayesian optimization toolkit, refining the geometry parameters.
Both steps are repeated alternatingly until the quality factor is maximized for the fundamental cavity mode at the desired frequency.
For more details, refer to \cite{Bopp2025}.

\section{Results and Discussion}
\subsection{Dinosaur Unit Cell and Cavity Design}
\begin{figure}[htbp]
	\centering
	\includegraphics[scale=.7]{\figpath 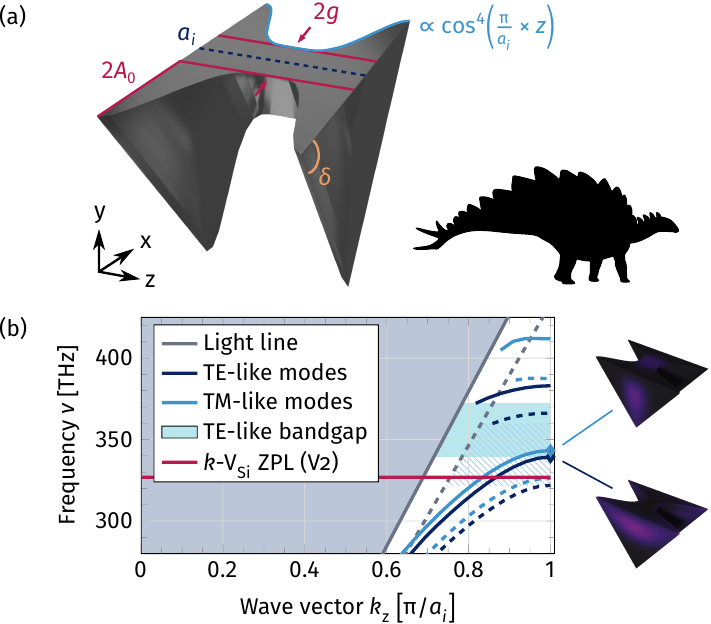}
	\caption{Dinosaur unit cell.
	(a) A Dinosaur unit cell of length $a_i$ features a triangular cross section, determined by the sidewall angle $\delta$, and a $\cos^4$ corrugation profile, which is defined by an amplitude $A_0$ and offset from the yz symmetry plane by the gap width $g$.
	It resembles the back plates of a Stegosaurus (inset).
	(b) A periodic sequence of identical unit cells yields an optical band diagram that supports TE-like (dark blue lines) and TM-like (light blue lines) Bloch bands for wave vectors $k_\mathrm{z}$ along the stacking direction that exceed the light line $\nu = c|\vec{k}|$ (gray line).
	$c$ is the speed of light.
	Solid lines refer to a unit cell length $a_0=315.9\,$nm, and dashed lines to a length $a_3=346.1\,$nm.
	A TE-like optical bandgap (cyan-shaded area for $a_0$, dashed area for $a_3$) separates the two lowermost TE-like Bloch bands.
	The \textit{k}-$\mathrm{V}_\mathrm{Si}$ ZPL emission frequency is indicated by a red solid line.
	Insets visualize the electric field intensities of both the fundamental TE-like (lower inset) and TM-like (upper inset) Bloch modes at $k_\mathrm{z}=\pi/a_0$ (blue diamonds).}
	\label{fig:unitcell}
\end{figure}

Dinosaur PhC nanostructures are composed of Dinosaur unit cells (Fig.~\ref{fig:unitcell}(a)), which roughly resemble the back plates of a Stegosaurus (inset).
The Dinosaur unit cell geometry is defined by the length $a_i$, the sidewall angle $\delta$, and a corrugation profile $x(z) = 2 A_0 \cos^e(\pi/a_i\times z) + g$ with an even exponent $e$.
The corrugation features are proportional to the corrugation amplitude $A_0$ and offset from the central yz symmetry plane by the gap width $g$.
We select $\delta=54^\circ$, constrained by the Faraday cages for angled etching of 4\textit{H}-SiC available to us.
A stack of identical unit cells arranged along the z-direction forms a periodic PhC structure that may guide light as Bloch modes.
However, light can escape the PhC structure for wave vectors permitted by the linear dispersion relation of plane electromagnetic waves in vacuum (gray-shaded area in Fig.~\ref{fig:unitcell}(b)).
Choosing the parameters $a_0=315.9\,$nm, $A_0=124.7\,$nm, $g=75.0\,$nm, and $e=4$ allows the Bloch mode propagation of the \textit{k}-$\mathrm{V}_\mathrm{Si}$ V2 ZPL emission, as the ZPL frequency (red line) intersects with the fundamental TE- and TM-like Bloch bands (lowermost blue solid lines).
These bands correspond to dielectric modes with an elliptical electric field intensity cross section.
For TE- and TM-like modes, the semi-major axis of their mode cross section is parallel to the x- and y-directions, respectively (insets in Fig.~\ref{fig:unitcell}(b)).
In between both displayed TE-like Bloch bands, there is a TE-like bandgap with a center frequency of approximately $355.7\,$THz and a gap-midgap ratio of approximately $9.2$\,\% (cyan-shaded are).
The selected exponent and gap width represent a tradeoff.
They preserve sufficient material between adjacent corrugation maxima for nanofabrication and open up a photonic bandgap large enough to enable high quality factors.
Increasing the unit cell length to $a_3=346.1\,$nm shifts all Bloch bands to lower frequencies (dashed lines).
Hence, the V2 ZPL emission falls into the TE-like bandgap of a periodic Dinosaur PhC nanostructure with unit cell lengths $a_3$ (dashed area), and is not allowed to propagate anymore.

\begin{figure}[htbp]
	\centering
	\includegraphics[scale=.7]{\figpath 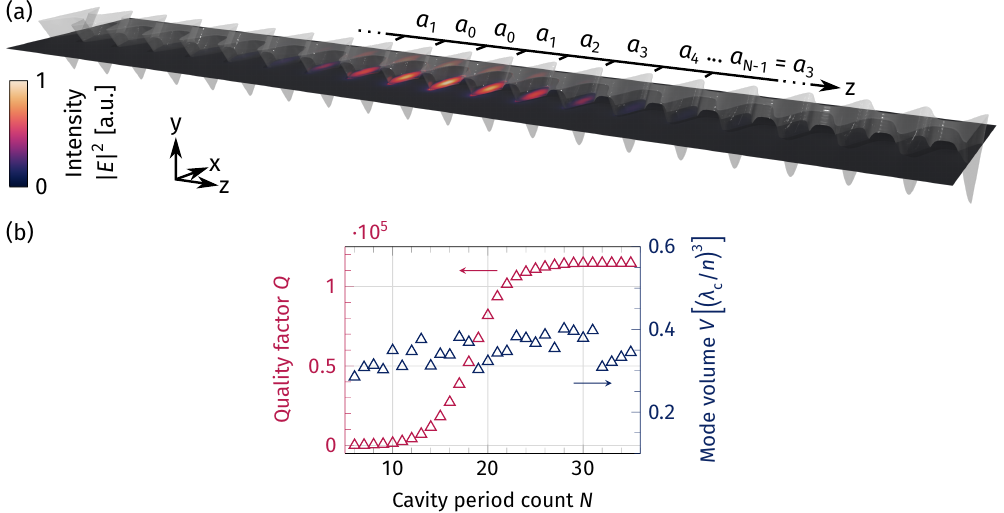}
	\caption{Symmetric Dinosaur cavity.
	(a) Starting with a stacked sequence of Dinosaur unit cells with a length $a_3$, a PhC cavity with an additional xy symmetry plane at $z=0$ is introduced by decreasing the length of the innermost three unit cells monotonically to $a_0$.
	The unit cell with length $a_0$ ends at $z=0$, where the confined cavity mode's electric field intensity is maximal, as visualized in an xz cross-sectional plane through the cavity.
	Along the y-direction, the plane is located at $y=0$, i.e., at a depth below the cavity surface matching one-third of the unit cells' triangular cross section height.
	(b) The quality factor $Q$ (red triangles) and the mode volume $V$ (blue triangles) of the confined cavity mode depends on the the number of stacked unit cells $N$, counting from the cavity's xy symmetry plane to one of its ends.
	Here, $\lambda_\mathrm{c}=917.0\,$nm denotes the cavity's resonance wavelength after convergence ($N\geq 24$), matching its resonance frequency $\nu_\mathrm{c}=326.9\,$THz, and $n$ the refractive index of 4\textit{H}-SiC (refer to main text).
	Panel (b) is adapted from \cite{Bopp2025} and originally published under CC BY 4.0.}
	\label{fig:cavity}
\end{figure}

Stacking unit cells with different lengths along the positive and negative z-direction, as shown in Fig.~\ref{fig:cavity}(a), constitutes a symmetric Dinosaur PhC cavity.
To confine light to the cavity center, the V2 ZPL emission has to be permitted in this region and forbidden in the outer cavity regions with a smooth transition in between to avoid scattering losses.
Consequently, both innermost unit cells interfacing the xy symmetry plane possess a length $a_0$.
Toward the cavity ends, the lengths $a_i$ of $N$ unit cells gradually increase in each symmetric half of the cavity via $a_1=321.9\,$nm and $a_2=332.8\,$nm to $a_3 \dots a_{N-1}=346.1\,$nm.
Fig.~\ref{fig:cavity}(a) additionally depicts the confined TE-like electric field intensity in an xz plane at a depth below the cavity surface of $148.8\,$nm that equals one-third of the maximal unit cell height in y-direction.
Cavity-embedded \textit{k}-$\mathrm{V}_\mathrm{Si}$ centers are supposed to be located within this plane at the cavity mode maximum for optimal coupling.
For cavity period counts $N\gtrsim 30$, a quality factor $Q=1.1\times 10^5$ is reached, while the mode volume $V=0.35(3)\,(\lambda_\mathrm{c}/n)^3$ is rather independent of the cavity period count (Fig.~\ref{fig:cavity}(b)).
Here, $\lambda_\mathrm{c}=917.0\,$nm is the cavity resonance wavelength and $n=2.6$ \cite{Wang2013} the approximate refractive index of 4\textit{H}-SiC, neglecting its birefringence.
The mode volume fluctuates slightly depending on $N$, possibly indicating an undesired coupling mechanism between TE- and TM-like cavity modes.
Such coupling may be suppressed by choosing a sidewall angle further away from $45^\circ$ to lift the degeneracy between TE- and TM-like modes.

\subsection{Waveguide-Coupled Dinosaur Cavity}
\begin{figure}[htbp]
	\centering
	\includegraphics[scale=.7]{\figpath 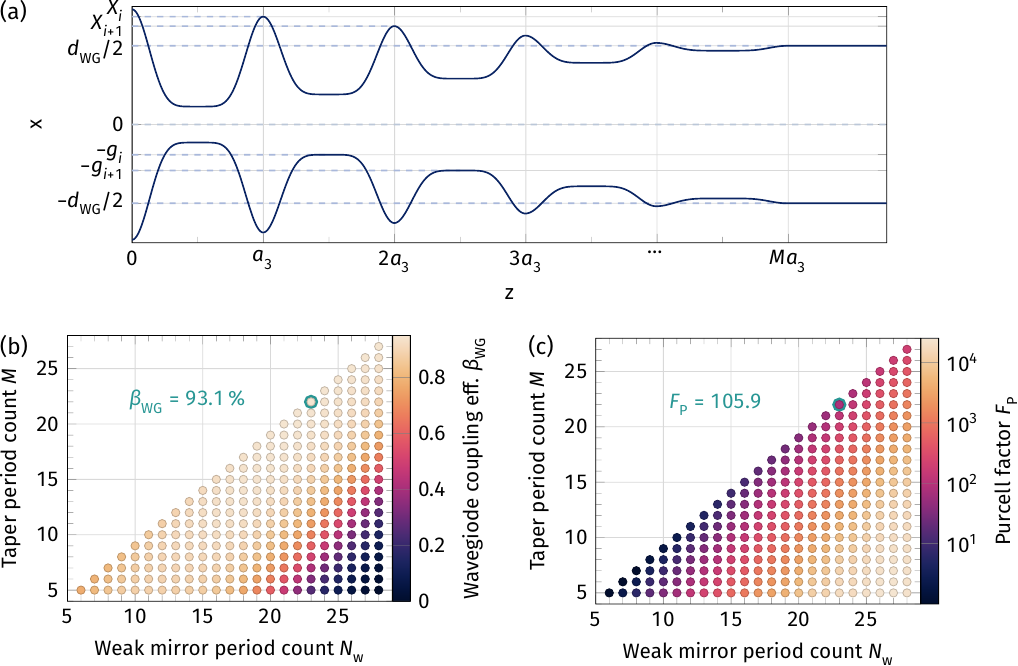}
	\caption{Tapered, asymmetric waveguide-coupled Dinosaur cavity.
	(a) The taper profile $x(z)$ of the tapered section interfacing the cavity with a waveguide of width $d_\mathrm{WG}$ is modulated by a third-order polynomial.
	Its peak heights $X_i$ and gap widths $g_i$ are determined by Eq.~\ref{eq:taper_A} and Eq.~\ref{eq:taper_P}, respectively.
	The profile is symmetric with respect to $x=0$.
	(b) The waveguide coupling efficiency $\beta_\mathrm{WG}$ depends on the number of unit cells $N_\mathrm{w}$ constituting the cavity's weak mirror and on the number of unit cells $M$ forming the tapered section within the weak mirror.
	(c) Likewise, the Purcell factor $F_\mathrm{P}$ depends on the weak mirror parameters $N_\mathrm{w}$ and $M$.
	In (b) and (c), the cavity's strong mirror possesses $N_\mathrm{s}=30$ unit cells.
	The encircled cavity configuration exhibits the highest observed waveguide coupling efficiency of $93.1\,\%$, corresponding to a Purcell factor of $105.9$.}
	\label{fig:taper}
\end{figure}

Efficiently interfacing the cavity-embedded color centers through a cavity-attached waveguide requires a tapered section between the cavity and the waveguide where the corrugation features fade, meaning that the corrugation amplitude has to decrease gradually (Fig.~\ref{fig:concept}).
This section adiabatically converts the waveguide to a Bloch mode and vice versa, minimizing scattering losses at the cavity--waveguide interface \cite{Palamaru2001}.
The width at the waveguide-attached end of the tapered section is defined as $d_\mathrm{WG}$.
It is chosen such that the waveguide does not support higher-order optical modes.
Tapered sections are incorporated into Dinosaur PhC cavities by modification of one of the two cavity mirrors, rendering the cavity asymmetric.
On one side, the cavity now has a strong mirror consisting of $N_\mathrm{s}$ unit cells.
A small fraction of light is outcoupled through a weaker mirror composed of $N_\mathrm{w}$ unit cells on the other side.
Within the weak mirror, the outermost $M$ unit cells form the tapered section.
For an adiabatic transfer of the cavity mode into a waveguide mode, we gradually reduce the corrugation amplitude while widening the central gap.
With $i\in[0,M]$, the taper profile is defined by the peak heights $X_i$ and the modified gap widths $g_i$ between the corrugation maxima (Fig.~\ref{fig:taper}(a)).
Both $X_i$ and $g_i$ are described by a third-order polynomial.
To obtain seamless transitions at the cavity--tapered section and tapered section--waveguide interfaces, we introduce the boundary conditions $\mathrm{d}/\mathrm{d}z(X_i) = \mathrm{d}/\mathrm{d}z(g_i) = 0$ at the beginning and end of the tapered section.
By matching the polynomial to the cavity parameters $A_0$, $g$, and the target waveguide width $d_{\textrm{WG}}$, the polynomial coefficients, and thus $X_i$ and $g_i$ that yield the taper profile $x(z)$, are uniquely determined by 
\begin{align}
	X_i &= \frac{d_\mathrm{WG}}{2} - 3\left(\frac{d_\mathrm{WG}}{2}-X_0\right)q^2 + 2\left(\frac{d_\mathrm{WG}}{2}-X_0\right)q^3 \,, \label{eq:taper_A}\\
	g_i &= \frac{d_\mathrm{WG}}{2} - 3\left(\frac{d_\mathrm{WG}}{2}-g\right)q^2 + 2\left(\frac{d_\mathrm{WG}}{2}-g\right)q^3 \,, \label{eq:taper_P}
\end{align}
where $q = (M-i)/M$ and $X_0 = 2A_0+g$.
Consequently, the profile of the $i$-th unit cell is given by 
\begin{align}
	x(z) = (X_i-g_i) \cos^4\!\left(\frac{\pi}{a_3}\times z \right) + g_i \,.
\end{align}
This approach provides flexibility, as the target waveguide width and the Dinosaur PhC cavity parameters can be adapted to the desired use case. 
The imposed boundary conditions uniquely determine the resulting taper profile, allowing the method to be applied to any set of cavity and waveguide parameters.

To investigate the influence of different tapered section geometries, scattering simulations are performed with dipoles oriented along the x-direction and positioned at the cavity center at a depth below the surface as indicated by the electric field intensity plane displayed in Fig.~\ref{fig:cavity}(a).
The dipole frequency is matched to the cavity resonance frequency that is determined beforehand for each cavity configuration by performing a respective eigenfrequency simulation.
The strong mirror period count is fixed to $N_\mathrm{s}=30$ unit cells since the intrinsic cavity quality factor has fully converged for such a value (Fig.~\ref{fig:cavity}(b)).
This choice minimizes the cavity decay rate through the strong mirror.
Fig.~\ref{fig:taper}(b) and Fig.~\ref{fig:taper}(c) display the dependence of the cavity--waveguide coupling efficiency $\beta_\mathrm{WG}$ and the Purcell factor $F_\mathrm{P}$, respectively, on the weak mirror parameters $N_\mathrm{w}$ and $M$, i.e., the length of the weak mirror and the length of the tapered section within.
Both parameters determine the decay rate of the cavity mode into the waveguide and into the surrounding environment but also the Purcell enhancement and thus the strength of the light-matter interaction.
For a broad range of weak mirror parameters, light is coupled efficiently into the waveguide.
On the one hand, only for `strong' weak mirrors with a large number of periods and a short tapered section, the waveguide coupling efficiency drops since cavities again approach the symmetric cavity limit in this regime.
In the symmetric cavity limit, light does not preferentially couple from the cavity mode to the waveguide mode but is rather scattered to the environment.
On the other hand, the Purcell factor is highest in the symmetric cavity limit where a cavity consists of two opposing strong mirrors that confine light longest to the cavity center due to a high quality factor.
However, the right choice of weak mirror parameters enables tuning the Purcell factor across orders of magnitude, while a near-unity waveguide coupling efficiency is maintained.
Exemplarily, for $N_\mathrm{w}=23$ and $M=22$ (encircled data points in Fig.~\ref{fig:taper}(b) and Fig.~\ref{fig:taper}(c)), we obtain the maximal waveguide coupling efficiency $\beta_\mathrm{WG}=93.1\,\%$ in conjunction with a Purcell factor $F_\mathrm{P}=105.9$.
Similar dependencies of $\beta_\mathrm{WG}$ and $F_\mathrm{P}$ on the weak mirror parameters have been observed for the Sawfish cavity \cite{Bopp2023a}.
Hence, we deem these dependencies universal to waveguide-coupled PhC cavities.
Since the Purcell factor relates to the cooperativity $C$ of the coupled atom--cavity system, with color centers acting as solid-state artificial atoms, tuning the weak mirror parameters may facilitate entering the $C\gg 1$ regime.
This regime is characterized by nonlinear interactions between the cavity-coupled atom and cavity mode photons \cite{Reiserer2015}, and is therefore of particular interest for quantum information technologies \cite{Nguyen2019}.

\subsection{Dinosaur Cavity for Diamond Color Centers}
\begin{figure}[htbp]
	\centering
	\includegraphics[scale=.7]{\figpath 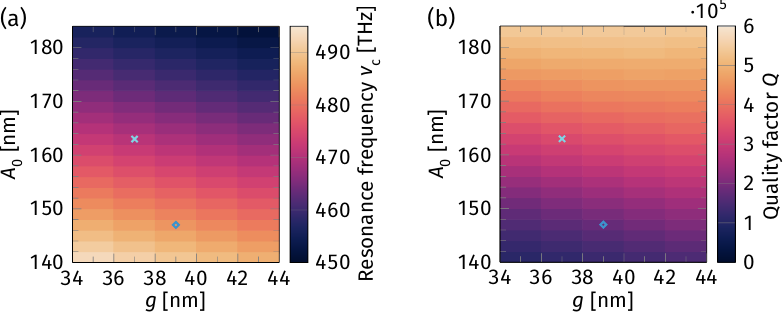}
	\caption{Cavity parameter adjustment for diamond color centers.
	(a) The resonance frequency $\nu_\mathrm{c}$ of Dinosaur PhC cavities can be tuned by varying the gap width $g$ or the corrugation amplitude $A_0$ to accommodate the ZPL of the negatively-charged diamond NV (cyan crosses) or SnV (blue diamonds) centers.
	(b) Such tuning likewise affects the quality factor $Q$.
	The figure is adapted from \cite{Bopp2025} and originally published under CC BY 4.0.}
	\label{fig:diamond}
\end{figure}

To adapt the symmetric Dinosaur PhC cavity to diamond color centers, the sidewall angle is decreased to $\delta=35^\circ$.
A reduced sidewall angle is assumed to alleviate the coupling between differently polarized cavity modes.
Due to the resulting triangular unit cell cross section, which is compressed along the y-direction, large parts of the fundamental TM-like Bloch cannot be guided inside the diamond nanostructure anymore.
This increases the energy splitting between both fundamental Bloch bands.
Moreover, the unit cell lengths become $a_0=209.5\,$nm, $a_1=213.2\,$nm, $a_2=220.8\,$nm, $a_3=227.9\,$nm, and $a_4 \dots a_{N-1}=238.6\,$nm, targeting a fundamental TE-like cavity mode frequency of about $471\,$THz.
Varying the additional lattice constant $a_4$ benefits cavity quality factors in case of diamond Dinosaur cavities but not the 4\textit{H}-SiC cavity introduced above.
Selecting the corrugation amplitude $A_0$ and the gap width $g$ according to Fig.~\ref{fig:diamond}(a) enables tuning the resonance frequency $\nu_\mathrm{c}$ of the fundamental TE-like cavity mode to the ZPL frequency of the negatively-charged NV and SnV centers at $470.8\,$THz (cyan cross) and $484.1\,$THz (blue diamond), respectively.
The corresponding quality factors $Q$ are $3.3\times 10^5$ for the NV and $1.8\times 10^5$ for the SnV center (Fig.~\ref{fig:diamond}(b)).
While the resonance frequency $\nu_\mathrm{c}$ rises with either decreasing gap width or corrugation amplitude, the quality factor diminishes.
Consequently, geometry parameter sweeps allow for efficient adaptation of the cavity resonance frequency to a desired color center ZPL frequency, at the cost of the optimized quality factor.
Dinosaur cavities for diamond color centers are adjusted to visible wavelengths, while the 4\textit{H}-SiC \textit{k}-$\mathrm{V}_\mathrm{Si}$ fluoresces in the near-infrared.
According to Fig.~\ref{fig:diamond}, lower quality factors would have been expected for cavity adaptations to diamond color centers compared to the \textit{k}-$\mathrm{V}_\mathrm{Si}$.
Advantageously, the quality factors do not decrease for visible wavelengths since the decreased sidewall angle avoids possible coupling between TE- and TM-like modes.
Here, excessive memory consumption of the diamond cavity simulations restricted the cavity period count to $N=14$.
Cavity period counts of $N>20$ are assumed to slightly improve the obtained quality factors.

\section{Conclusion and Outlook}
We introduce and investigate the Dinosaur PhC cavity, which features a complex corrugation profile and a triangular cross section.
In simulations, the cavity provides intrinsic quality factors in the order of $10^5$, a mode volume below $0.4\,(\lambda_\mathrm{c}/n)^3$, and a maximal cavity--waveguide coupling efficiency of $93.1\,\%$.
Both the light-matter interaction and the waveguide coupling efficiency are strongly affected by the cavity's tapered cavity--waveguide interface.
Hence, designing the tapered section of a PhC cavity allows for engineering the light-matter interaction itself \cite{Bopp2025}.
Furthermore, we adapt the cavity to various color center types in 4\textit{H}-SiC and diamond and showcase the respective geometry parameters.
These parameters can be tuned to increase the photonic bandgap or to adjust the sidewall angle to a certain Faraday cage geometry.

Compared to the Sawfish cavity with its rectangular cross section \cite{Bopp2023a, Pregnolato2024}, quality factors obtained for 4\textit{H}-SiC as well as diamond Dinosaur cavities are reduced by about one order of magnitude, indicating cross-talk between cavity modes with different polarization.
The mode volume of Dinosaur cavities reduces to less than half the Sawfish cavity mode volume since triangular cross sections squeeze the cavity mode into elliptical profiles.
In contrast to Sawfish cavities, Dinosaur cavities do not require quasi-isotropic undercutting but rely on Faraday etching, which necessitates fewer fabrication steps.

In next steps, Dinosaur cavities are to be fabricated and experimentally examined to measure their experimental quality factors and cavity--waveguide coupling efficiencies.
Moreover, hole-based triangular optomechanical crystal nanostructures exhibit photonic and phononic resonances at the same time \cite{Burek2016}, which enables acoustic interfacing of the spin degree of freedom of embedded color centers \cite{Raniwala2025}.
Exploring the Dinosaur cavity's mechanical modes is therefore an interesting avenue for future investigations.
In summary, the Dinosaur PhC cavity represents a promising building block for quantum photonic integrated circuits of repeater nodes in quantum communication networks since it can realize efficient spin--photon interfaces with relaxed demands on nanofabrication.

\paragraph*{Funding}
This project is funded by the German Federal Ministry of Research, Technology and Space (BMFTR) within the `QPIC-1' project (No. 13N15858), the `QR.N' project (No. 16KIS2185), and the `DIAQUAM' project (No. 13N16957) as well as by the European Research Council within the ERC Starting Grant `QUREP' (No. 851810).

\paragraph*{Acknowledgment}
The authors thank Matthias Plock and Sven Burger for advice on performing efficient simulations with JCMsuite, as well as Tommaso Pregnolato for fruitful discussions about the tapered section's mathematical definition.

\paragraph*{Disclosures}
The authors declare no conflicts of interest.

\paragraph*{Data Availability Statement}
Simulation data underlying the results presented in this paper are not publicly available at this time but may be obtained from the authors upon reasonable request. No experimental data were generated or analyzed in the presented research.

\bibliographystyle{apsrev4-2}
\bibliography{dinosaur.bib}
\end{document}